\documentclass{aa}  
\usepackage{ragged2e}

\usepackage{float}
\usepackage{graphicx}
\usepackage{tablefootnote}
\usepackage{hyperref}
\usepackage{academicons}
\usepackage{tikz,xcolor,hyperref}

\newcommand{\orcidauthorA}{0009-0009-8988-0537}
\newcommand{\orcidauthorB}{0000-0001-6264-140X}
\newcommand{\orcidauthorC}{0000-0003-0079-1239}
\newcommand{\orcidauthorD}{0000-0002-0273-218X}

\usepackage{orcidlink}

\hypersetup{
    colorlinks=true,
    citecolor=blue,
    linkcolor=blue,
    urlcolor=blue,
}

\usepackage{xcolor}
\usepackage{natbib}
\usepackage{txfonts}
\usepackage{comment}

\def\medd{$\dot{m}_{\rm Edd}$}
\def\mbh{$M_{\rm BH}$}
\def\psif{$\Psi_{\rm \lambda}(t)$} 
\def\gammaf{$\Gamma_{\rm \lambda}(\nu)$}

\def\spin{$\alpha^{*}$}
\def\gammaff{$|\Gamma_{\rm \lambda}(\nu)|^{2}$}
\def\ltransf{$L_{\rm transf}/L_{\rm disc}$}

\def\gammanorm{$|\Gamma_{\rm \lambda, norm}(\nu)|^{2}$}
\def\psdX{$\textrm{PSD}_{\rm X}(\nu)$ }
\def\fkynsed{$F_{\lambda, \rm KYNSED}$}

\def\fcol{$f_{\rm col}$}
\def\Rout{$R_{\rm out}$}
\def\rtransf{$r_{\rm transf}$}
\def\psilt{$\Psi_{\rm \lambda}(t)$}

\begin{document} 

\title{X-ray reverberation modelling of the observed UV/optical power spectra of quasars}

\author{M. Papoutsis \inst{1,2}
\and
I. E. Papadakis\inst{3,2}
\and
C. Panagiotou\inst{4} 
\and 
E. Kammoun\inst{5}
\and
M. Dovčiak\inst{6}
}

\titlerunning{UV/opt AGN variability}
\authorrunning{M. Papoutsis et al.}

\institute{Department of Physics, University of Crete, 71003 Heraklion, Greece \email{mpapoutsis@physics.uoc.gr}
\and 
Institute of Astrophysics, FORTH, GR-71110 Heraklion, Greece
\and
Department of Physics and Institute of Theoretical and Computational Physics, University of Crete, 71003 Heraklion, Greece
\and
MIT Kavli Institute for Astrophysics and Space Research, Massachusetts Institute of Technology, Cambridge, MA 02139, USA
\and
Cahill Center for Astronomy and Astrophysics, California Institute of Technology, 1200 California Boulevard, Pasadena, CA 91125, USA
\and
Astronomical Institute of the Academy of Sciences, Boční II 1401, CZ-14100 Prague, Czech Republic
}


   \date{Received XXXX; accepted YYYY}

 
  \abstract
{Over the past decade, a significant amount of effort has been put into investigating the ultraviolet (UV) and optical variability of active galactic nuclei (AGNs). Comprehensive studies of intensive multi-wavelength monitoring and surveys of local and high-redshift AGNs have shown that X-ray illumination of AGN accretion discs is a potential explanation for the observed variability.}
{Our main objective is to study the UV/optical power spectra of AGNs under the assumption of X-ray reverberation and to test whether this model can explain the observed power spectra of distant quasars.}
{We computed the disc transfer function in the case of X-ray reverberation using a recent physical model and studied its dependence on the parameters of the model. This model allows us to explore the variability of X-ray illuminated discs under the scenario in which the X-ray corona is powered by the accretion process
or by an external source. We then calculated UV/optical power spectra using the disc transfer function and assuming a bending power law for the X-ray power spectrum. We fitted our models to the observed power spectra of quasars determined by a recent power spectrum analysis of the SDSS Stripe-82 light curves.}
{We demonstrate that X-ray reverberation can fit the power spectra of quasars in our sample well at all wavelengths, from $\sim 1300$\AA\ up to $4000$\AA. Our best-fit models imply that the X-ray corona is powered by the accretion disc, and that the black hole spin is probably lower than 0.7, while the X-ray corona height is in the range of $20 - 60 R_{g}$. This is in agreement with previous findings from the application of the X-ray reverberation model to the quasar micro-lensing disc size problem, as well as recent time-lag measurements. }
   {}
   

   \maketitle
%
\section{Introduction}

It is generally accepted that active galactic nuclei (AGNs) are powered by the accretion of matter onto a supermassive black hole (SMBH) in the form of a geometrically thin and optically thick disc \cite[NT73 hereafter]{SS73, NT73}. The bulk of the disc radiation is emitted in the ultraviolet and optical parts of the electromagnetic spectrum (UV/optical hereafter). Active galactic nuclei are also known to be strong X-ray emitters. It is believed that the X-rays are produced as disc photons are upscattered by high-energy electrons residing in the so-called X-ray corona. However, the exact geometry and nature of the corona remain unknown. 

A defining characteristic of an AGN is flux variability, observed across all wavebands but with varying amplitudes and timescales. Typically, the amplitude of variability increases, while the characteristic timescales decrease as the observing wavelength decreases \citep[see e.g.][for a review]{Ulrich97}. Investigating these flux variations offers valuable insights into the physical processes occurring in the innermost regions of active galaxies.

In addition to recent intensive, multi-wavelength campaigns of nearby Seyferts \citep[e.g.][]{Cackett2020, Kara21, Edelson24}
, the last two decades have also seen an increase in both the quality and quantity of multi-epoch UV/optical data of quasars from large ground-based surveys \citep[e.g.][]{Bauer09, McLeod10, Zu2013, dechico22}. Space-based missions such as {\it Kepler} and the {\it Transiting Exoplanet Survey Satellite}  (TESS) have also provided data with a high cadence, although for a smaller number of objects \citep[e.g.][]{Mushotzky11, Smith18, Treiber23}. 

Recently, \citet[][P24 hereafter]{Petrecca24} analysed archival data from the SDSS Stripe-82 region to directly compute the power spectrum of quasars in the (rest frame) UV/optical bands. They fitted the observed power spectra with phenomenological models and used the results to study how the quasar power spectral density (PSD) depends on the black hole (BH) mass (\mbh), bolometric luminosity, accretion rate, redshift, and rest-frame wavelength. They also discussed the possibility of a universal power spectral shape for all quasars, whereby frequencies scale with the BH mass, while normalisation and slope(s) are fixed (at any given wavelength and accretion rate).

Our main objective is to study the power spectra of AGNs in the UV/optical bands, assuming that all observed variations are due to X-ray reverberation, and to investigate whether X-ray reverberation can explain the UV/optical observed power spectra of quasars, determined by P24. The hypothesis of disc X-ray illumination originated many years ago, after the detection of the Fe K$\alpha$ emission line around 6.4 keV and the Compton hump in the X-ray spectra of AGNs \cite[e.g][]{Pounds90}. This hypothesis makes certain predictions regarding the UV/optical variability of AGNs. First, if the X-rays illuminate the disc and are variable, then the disc emission should be variable in a correlated way, but with a delay that should be representative of the light travel time from the X–ray corona to the disc. Such X-ray/optical correlations and delays have already been observed in a few Seyert galaxies, and also in quasars using light curves from long-term monitoring campaigns \citep[e.g.][]{Jiang_17, Mudd_18, Homayouni_19, Yu_20, Jha22}. Recently, \citet{Langis24} showed that the observed time lags can be explained within the context of X-ray reverberation. 

Secondly, if the X-ray PSD is known, then one can also predict the UV/optical power spectrum of a source under the hypothesis of X-ray illumination of the disc, as the UV/optical PSDs depend only on the main physical characteristics of the system (i.e. BH mass and spin, accretion rate, and the X-ray corona luminosity and height). \citet[][P22 hereafter]{Panagiotou222} showed that the UV/optical PSD of NGC 5548 can indeed be modelled well by assuming X-ray reverberation, given the observed X-ray PSD of the source. In a similar work, \cite{Papoutsis24} showed that the observed UV/optical variance in many wavelengths of a few AGNs is fully consistent with the predictions of X-ray reverberation, given the observed X-ray variations in the same objects. 

This work is divided into two parts. The first part is similar to the work of P22: we compute the disc transfer function and study its dependence on the various physical parameters. To do this, we implement the updated code from \citet[][K23 hereafter]{Kammoun23}, which allows us to explore the variability of X-ray illuminated discs under the scenario in which the accretion process powers the X-ray corona. In the second part of this work, we use the disc-transfer functions to compute model power spectra. We then fit our physical power spectra models to the ensemble power spectra of SDSS quasars calculated in P24 to test whether X-ray reverberation can explain the UV/optical variability of the quasars, as well as the ‘universal’ PSD shape for quasars that P24 discussed in their paper. 

This work is organised as follows. In Sect.\,\ref{sec:rev_model} we describe the set-up of the X-ray reverberation model. In Sect.\,\ref{sec:transfer} we briefly explain how the new code works and how we calculated the disc transfer function, and investigate its dependence on the various model parameters. In Sect.\,\ref{sec:obs} we describe the quasar sample of P24 that we use in this work. In Sect.\,\ref{sec:psd} we explain the calculation of the model power spectra and in Sect.\,\ref{sec:fitting} the fitting procedure. We discuss the best-fit results in Sect.\,\ref{sec:best_fit} and summarise our work in Sect.\,\ref{sec:discuss}. 

\section{The X-ray reverberation model}
\label{sec:rev_model}

Consider an X-ray source that illuminates an accretion disc. Part of the X-ray emission falling on the disc will be re-emitted in X-rays (this is known as the `disc reflection component'). The other part will be absorbed by the disc and act as an additional source of heating. In this way, the disc emission is connected with the X-ray emission. This connection can be expressed as
\begin{equation}
    F_{\rm \lambda}(t) = F_{\rm \lambda, NT} +
     \int_{0}^{\infty} \Psi_{\lambda}(t')F_{X}(t-t')dt' \;,
\label{eq:discflux}
\end{equation}
where $F_{\rm \lambda}(t)$ is the total disc flux at wavelength $\lambda$ and time $t$, $F_{\rm \lambda, NT}$ is the constant flux of a standard accretion disc when there is no illumination (NT73), $F_{X}(t)$ is a quantity representative of the X-ray flux at time $t$, and $\Psi_{\lambda}(t)$ is the disc response function at time $t$. 
$\Psi_{\lambda}(t)$ describes how the disc responds to the illumination of X-rays \citep[see][K21a hereafter]{Kammoun211}.

Equation\,(\ref{eq:discflux}) demonstrates that the emission of the disc at any given time, $t$, depends on the X-rays emitted in the past, with the details of this relationship captured by the form of the response function. The shape of the response function depends on all the physical parameters of the accretion disc/X-ray source system such as the BH mass, the accretion rate, and the geometry and luminosity of the X-ray source, among others. As a result, comparing the observed disc emission with the one calculated by Eq.\,(\ref{eq:discflux}), for given X-ray emission, and different physical models of the response function, can provide us with information about the AGN accretion process. However, working in the time domain is challenging due to the complex nature of the convolution integral in Eq.\,(\ref{eq:discflux}) and the rapid variability of the X-ray source. 

This analysis is simplified by working in the frequency domain. If the X-rays are variable, which they are in AGNs, then we also expect the disc emission, at all wavelengths, to be variable. In this case, if Eq.\,(\ref{eq:discflux}) holds, the power spectrum at wavelength $\lambda$, $\textrm{PSD}_{\lambda}(\nu)$, will depend on the X-ray power spectrum, $\textrm{PSD}_{X}(\nu)$, as follows \citep[e.g.][]{Papadakis16,Panagiotou222}:

\begin{equation}
    \textrm{PSD}_{\rm \lambda}(\nu) = |\Gamma_{\rm \lambda}(\nu)|^{2} \cdot \textrm{PSD}_{\rm X}(\nu) \; ,
\label{eq:psds}
\end{equation}
\noindent where $\Gamma_{\lambda}(\nu)$ is the disc transfer function, which is equal to the Fourier transform of the response function; that is,
\begin{equation}
\Gamma_{\lambda}(\nu) = \int_{-\infty}^{\infty} \Psi_{\lambda}(t) e^{-2 \pi i \nu t} dt \; .
\label{eq:fourier}
\end{equation}
Just like \psif, \gammaff\ depends on the geometric and physical properties of the X-ray source and the accretion disc. For this reason, Eq.\,(\ref{eq:psds}) can also be used to constrain the physical properties of an AGN. Working in the Fourier space is a significant simplification, turning a convolution computation into a multiplication one. 

Our initial objective in this work is to study how the disc transfer function depends on the system parameters. Then, as an application,  we fit the observed UV/optical power spectra of quasars, determined recently by \cite{Petrecca24}. To this end, we calculate model UV/optical power spectra using Eq.\,(\ref{eq:psds}). We use Eq.\,(\ref{eq:fourier}) to calculate \gammaff\ for many physical parameters and we use empirical formulas found in the literature to compute $\textrm{PSD}_{X}(\nu)$.
We then fit the model UV/optical power spectra to the observed ones to investigate whether the disc X-ray reverberation model can explain the UV/optical variability of quasars.

\section{The disc transfer function}
\label{sec:transfer}

In this section, we compute disc transfer functions using Eq.\,(\ref{eq:fourier}) and study the shape and dependence of these functions on various system parameters. Regarding the disc response function \psilt, we calculate it using the new code of \cite{Kammoun23}, \texttt{KYNXiltr}, which we briefly describe below.

\subsection{Model assumptions and the disc response function}
\label{subsec:response}

We assume a Keplerian disc, extending from the innermost stable circular orbit (ISCO) to an outer radius, $R_{\rm out}$. The disc is illuminated by a point-like X-ray corona located on the axis above the BH at height $h$ (known as the `lamp-post' geometry). The X-ray source emits isotropically (in its rest frame) a spectrum of the form $F_X(t)=N(t) \rm E^{-\Gamma} \rm{exp}\left(-E/E_{cut}\right)$,  where $\Gamma$ and $\rm E_{cut}$ are assumed to be constant. All general relativity effects of light traveling from the corona to the accretion disc and from the disc and the corona to the observer are considered. Some of the X-rays reaching the disc are reflected, while the rest are absorbed and increase the disc temperature. The X-ray reflection spectrum is taken from the {\tt XILLVER} tables \cite[]{Garcia13, garcia_16}. We provide a short description of how \psilt\ is calculated below (see K21a, K23 for more details).

The code assumes an X-ray flash of temporal width $T_{\textrm{flash}} = 10t_{\rm g}$, with $t_{\rm g}=R_{\rm g}/c$\footnote{$R_{\rm g} = G M_{\rm BH}/c^{2}$, is the gravitational radius.}. The code computes the X-ray flux that is absorbed in each disc element as a function of time, identifies all the disc elements that brighten up at the same time (as seen by a distant observer) and calculates their total flux. The response function is equal to this flux minus their intrinsic flux given by NT73; that is, it is equal to the flux that the disc emits due to X-ray heating only (see Eq.3 in K21a). The disc flux is modelled by a colour-corrected blackbody with a spectral hardening factor \fcol. Following K23, the response is divided by the total X-ray luminosity of the flash, in Eddington units, and the duration of the flash. In this way, we computed $\Psi_{\lambda}$ in units of erg/s$^2$/cm$^2$.

The response function (and the transfer function) depends on the mass of the central BH, $M_{\rm BH}$, the accretion rate, \medd\ (measured in Eddington units throughout this work), the height of the X-ray corona, $h$, the X-ray luminosity, $L_{X}$, the X-ray spectrum photon index, $\Gamma$, the disc inclination, $\theta$, the colour correction factor, \fcol, the disc outer radius, \Rout, and the BH spin, \spin. X-ray luminosity is parametrised as the ratio of the total X-ray luminosity to the accretion power of the disc, $L_{\rm transf}/L_{\rm disc}$. When $L_{\rm transf}/L_{\rm disc}$ is positive (case A), $L_{X}$ is equal to the power released by the accretion process in the disc below a radius, $r_{\rm transf}$. This power is extracted from the disc and is assumed to power the corona through some unknown mechanism. When \ltransf\ is negative (case B), the power of the corona originates from an external source instead of the disc. 

In the next section, we show how we compute the disc transfer function \gammaf. We note that P22 have already computed the disc transfer function in the case of X-ray thermal reverberation. The main difference between P22 and our work is that P22 used the disc response functions of K21a to compute \gammaf. These responses were calculated for two values of BH spin (0, 1) and a fixed colour-correction factor of \fcol= 2.4. Additionally, their model assumed that the X-ray corona is powered by an external energy source, independent of the accretion process. In this work, we used the updated code of K23, which can compute \psilt\ for any BH spin, colour-correction factor, and in the case in which the X-ray corona is powered by the accretion process. P22 also studied the dependence of \gammaff\ on $M_{\rm BH}$, \medd, $L_{X}$, $h$, $\Gamma$, and  $\theta$. In the following, we compute the disc transfer function and show how it depends on the rest of the physical parameters; that is, \ltransf, \fcol, and \spin. 

\subsection{Computation of the disc transfer function}
\label{subsec:transfer_computation}

The disc response function, $\Psi_{\lambda}(t_k)$, in a given waveband with central wavelength $\lambda$ and width $\Delta \lambda$ was calculated at discrete time points, $t_{k}=k \cdot \Delta t$, where $k=0, 1, ..., N-1$, with $N$ being the total number of points and $\Delta t$ the size of the bin. Then, following Eq.\,(\ref{eq:fourier}), we computed the discrete Fourier transform as follows (see also P22):
\begin{equation}
\Gamma_{\lambda}(\nu_{j}) = \frac{\sum^{N-1}_{k=0}\Psi_{\lambda}(t_{k})e^{-2 \pi i \nu_{j} t_{k}} \Delta t}{\Delta \lambda \cdot e^{i \pi \nu_{j} T_{\textrm{flash}}} \textrm{sinc}(\pi \nu_{j}T_{\textrm{flash}})} \;.
\label{eq:fourier_discrete}
\end{equation}
The transfer function was
calculated at frequencies of $\nu_{j} = j/N \Delta t$, where $j = 1, 2, ..., N/2$, and has units of erg/s/cm$^2$/\AA. The division by $e^{i \pi \nu_{j} T_{\textrm{flash}}} \textrm{sinc}(\pi \nu_{j}T_{\textrm{flash}})$ accounts for the finite width of the X-ray flash \cite[see discussion in Appendix C of][]{Epitropakis16}.

We can use the disc transfer function to compute the UV/optical PSD model, following Eq.\,(\ref{eq:psds}). However, when one estimates the power spectrum from observed light curves, it is customary to normalise it to the mean of the light curve, in which case the PSD is an estimate of the variability amplitude normalised to the mean flux. For this reason, it is necessary to compute the normalised transfer function as well; for example, \gammanorm. To do this, we employed {\tt KYNSED} \cite[][]{Dovciak22}, which is a physical model for the broadband spectral energy distribution (SED) of X-ray illuminated accretion discs. Using {\tt KYNSED}, we calculated the flux of the disc at wavelength $\lambda$, \fkynsed, and computed the normalised disc transfer function as 

\begin{equation}
    |\Gamma_{\rm \lambda, norm}(\nu)|^{2} = |\Gamma_{\rm \lambda}(\nu)|^{2} / F^2_{\rm \lambda, KYNSED}.
    \label{eq:gnorm}
\end{equation}

\subsection{The shape of the disc transfer function}
\label{subsec:gamma2_shape}

When the corona is powered by an external source (case B), we find that \gammaff\ is constant at low frequencies and its amplitude depends on the amplitude of the response function. The transfer function remains constant up to a characteristic frequency, known as the break frequency, which depends on the width of the response function. At higher frequencies, \gammaff\ decreases rapidly following a shape similar to a power law. Our results are in agreement with the results of P22. 

When the corona is powered by extracting energy from the inner parts of the accretion disc (case A), the shape of \gammaff\ is the same as before at low frequencies. However, at higher frequencies, we notice a second frequency break. In this case, the slope of the transfer function becomes flatter, indicating a small excess of power at these frequencies. This behaviour is due to the reprocessed emission from the innermost regions of the disc, which causes a sharp peak in the response function at early times (see Fig.\,1 in K23). 

\begin{figure}
   \centering
    \includegraphics[width=9cm, height=6cm]{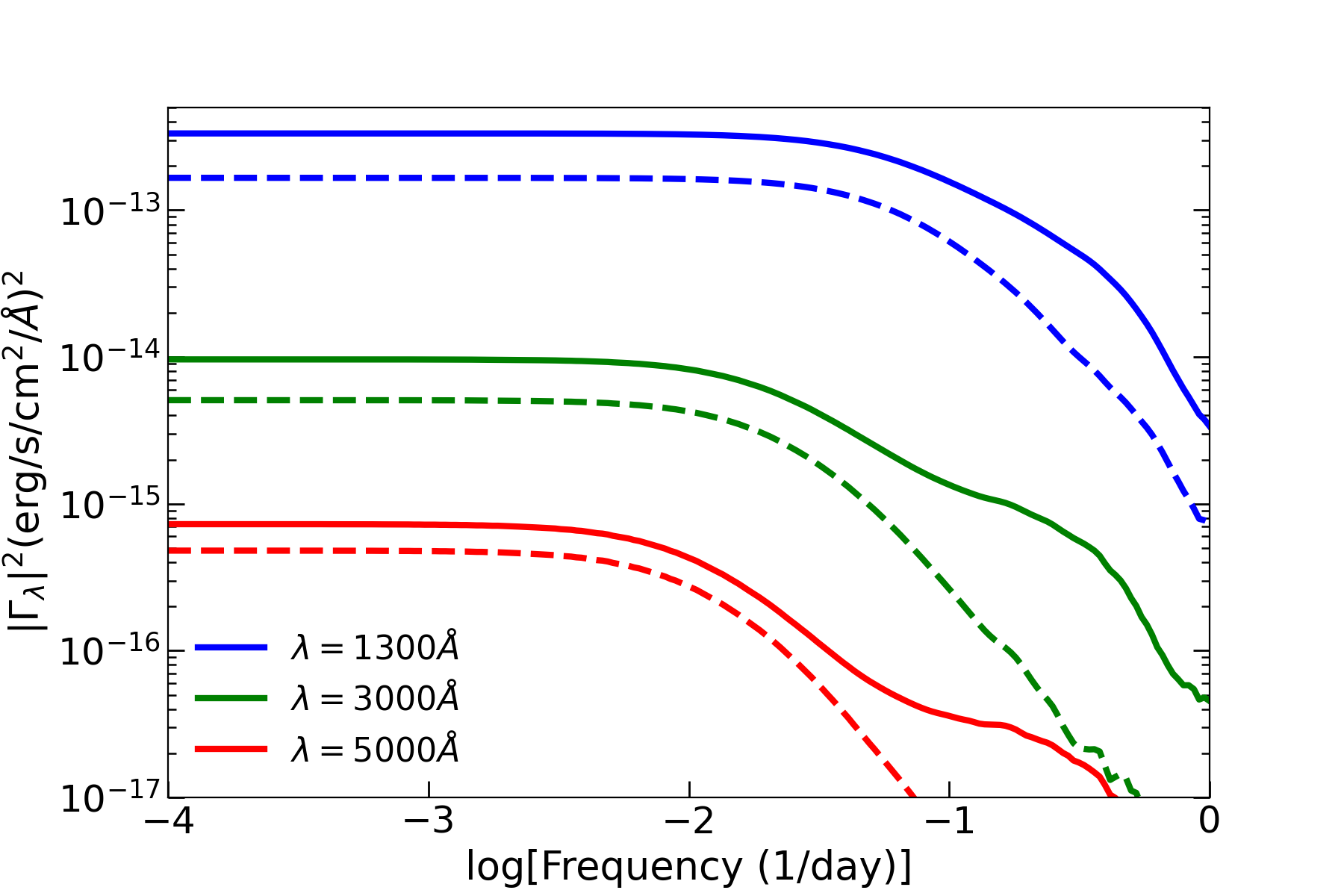}
    \includegraphics[width=9cm, height=6cm]{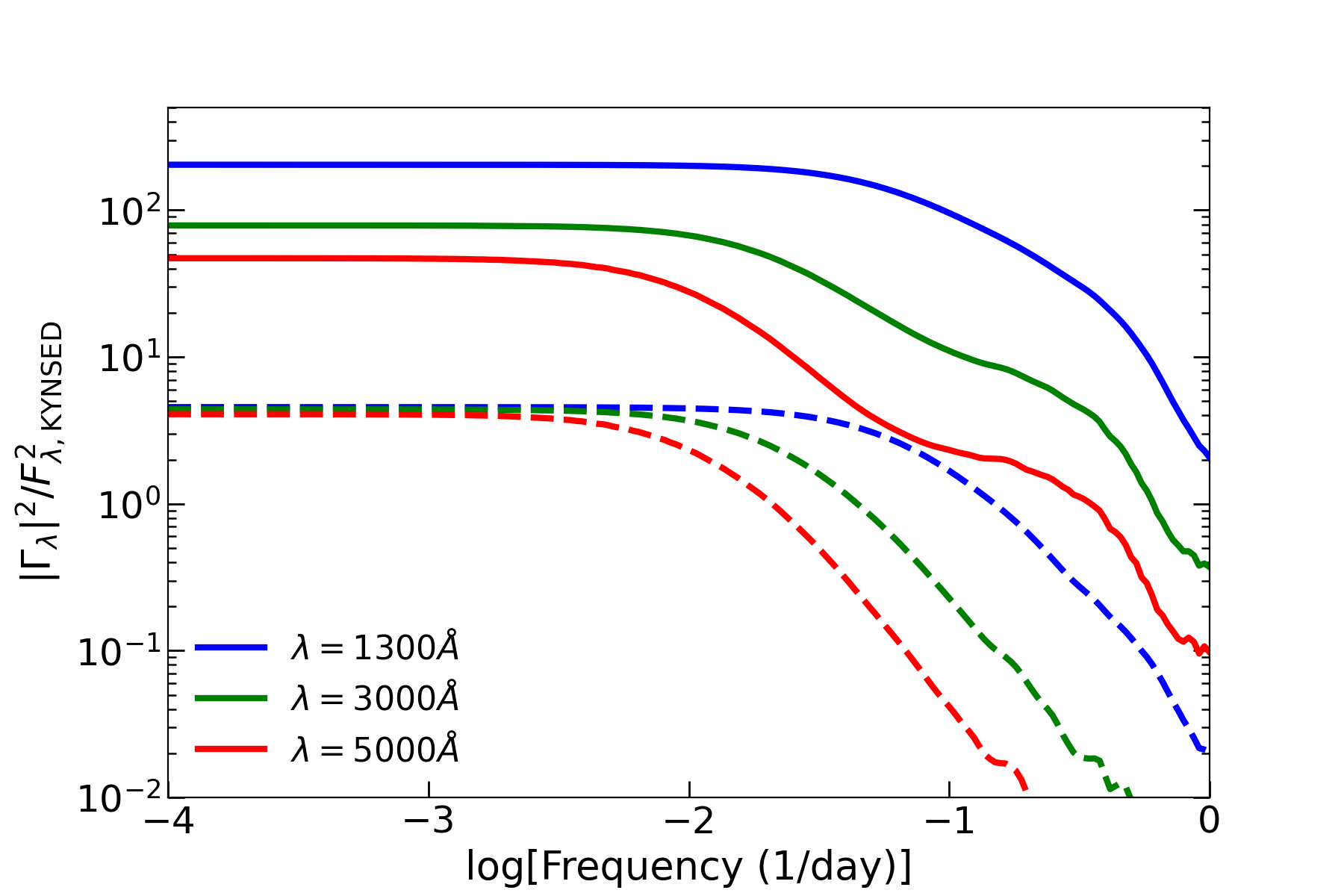}
      \caption{Transfer functions \gammaff\ (top panel) and normalised transfer functions \gammanorm\ (bottom panel) in three wavebands: $\lambda=1300$\AA\ (blue line), $\lambda=3000$\AA\ (green line), and $\lambda=5000$\AA\ (red line). The solid lines correspond to the case of an accretion-powered corona (case A) with $L_{\rm transf}/L_{\rm disc}=0.5$
     and the dashed lines to an externally powered corona (case B) with $L_{\rm transf}/L_{\rm disc}=-0.5$. The fiducial parameters are: $M_{\rm BH}=8 \times 10^8 M_{\odot}$, \medd=0.1, $h=20$, $\Gamma=2$, $\theta = 30^o$, $E_{\rm cut} = 150$keV, $f_{\rm col}=1.7$, $R_{\rm out}= 5000 R_{\rm g}$, and $\alpha^* = 0$.}
    \label{fig:gamma2_wave}
\end{figure}

\begin{figure*}
   \centering
    \includegraphics[width=19.cm, height=6cm]{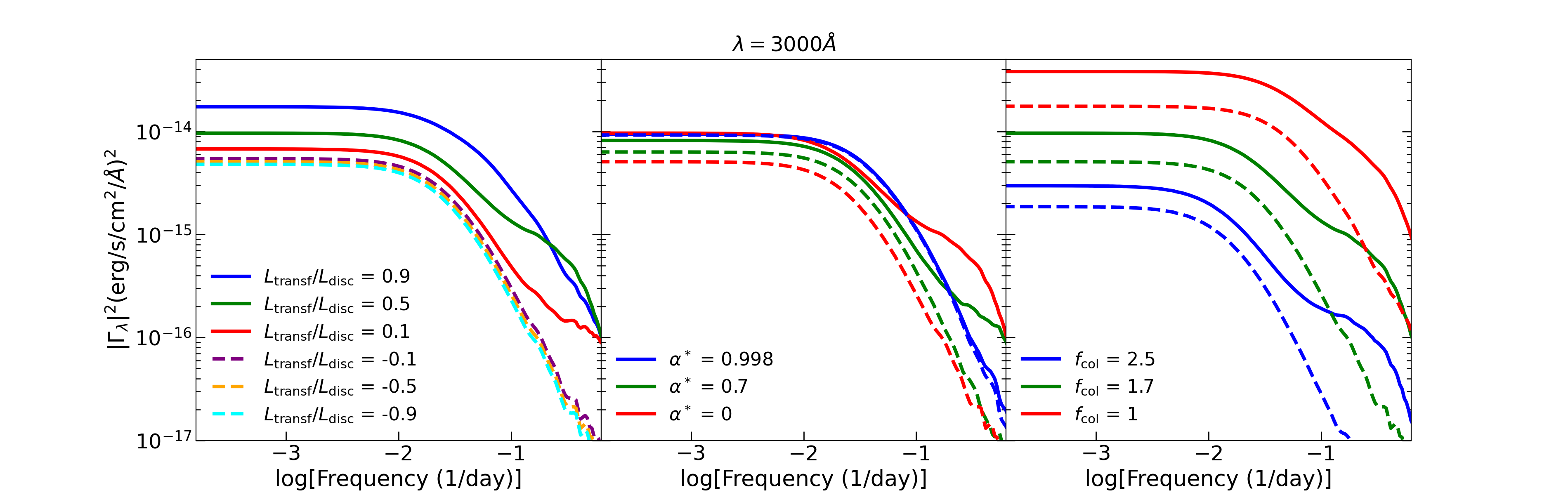}
    \includegraphics[width=19.cm, height=6cm]{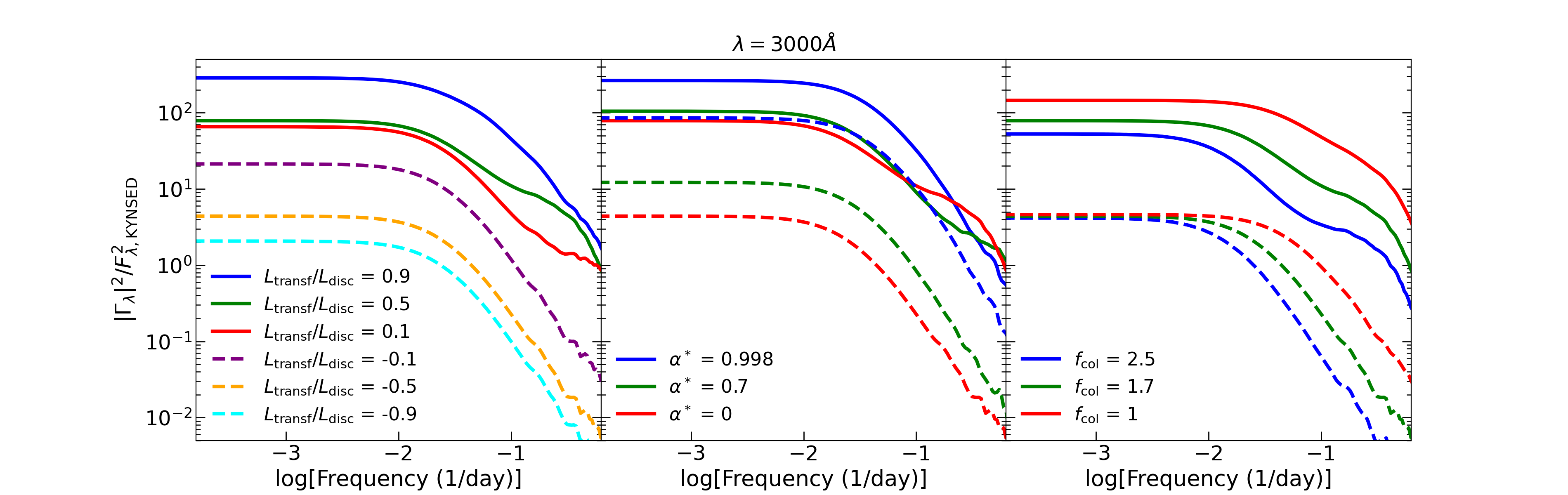}
      \caption{Transfer functions \gammaff\ (top panels) and normalised transfer functions \gammanorm\ (bottom panels) for different values of $L_{\rm transf}/L_{\rm disc}$ (left panels), \spin\ (middle panels), and \fcol\ (right panels)  at $\lambda=3000$\AA.  We show positive values of $L_{\rm transf}/L_{\rm disc}$ with solid lines (case A) and negative ones with dashed lines (case B). The fiducial parameters are: $M_{\rm BH}=8 \times 10^8 M_{\odot}$, \medd=0.1, $h=20$, $\Gamma=2$, $\theta = 30^o$, $E_{\rm cut} = 150$keV, $f_{\rm col}=1.7$, $R_{\rm out}= 5000 R_{\rm g}$, $\alpha^* = 0$, and \ltransf=-0.5, 0.5.}
    \label{fig:gamma2_params}
\end{figure*}

This feature in the transfer function (hence in the observed PSDs as well) can provide an insight into the source of power of the corona. It is a direct prediction of the model that can potentially help us differentiate between the
two scenarios for the origin of the X-ray corona power in AGNs. However, this feature 
appears at high frequencies. For example, it can be seen at frequencies of $\nu \gtrsim 0.1$ day$^{-1}$ in Figures \ref{fig:gamma2_wave}, \ref{fig:gamma2_params}, in which we have assumed a BH mass of $8\times10^8 M_\odot$. The observed PSDs of such sources (and of AGNs with higher-mass BHs) are probably dominated by the Poisson noise component at such frequencies. This is also the case for smaller BH-mass AGNs as well because this feature will move to even higher frequencies in these sources. Nevertheless, this is an important feature to look for in the observed PSDs of AGNs, and we plan to investigate this in future work.

P22 studied how the transfer function depends on the BH mass, the accretion rate, the height of the corona, and the inclination angle, in the case B corona. We found similar results even when the corona is powered by the accretion process (i.e. in the case A corona). In this work, we present results on how the transfer function changes with \ltransf, \fcol, and \spin, both in case A and case B coronae. 
In computing the transfer functions, we assume the following fiducial values for the system parameters: $M_{\rm BH}=8 \times 10^8 M_{\odot}$, $\dot{m}_{\rm Edd}=0.1$, $h=20$, $\Gamma=2$, $\theta = 30^o$, $E_{\rm cut} = 150$keV, and $R_{\rm out}=5000R_{g}$. For the parameters of interest, we assume \ltransf\ = -0.5, 0.5, \fcol\ = 1.7, and \spin\ = 0, when they are constant.

\subsection{The amplitude of the disc transfer function}

The amplitude of \gammaff\ depends on the 
overall shape of the response function. As is described in Sect.\, \ref{subsec:response}, \psif\ is proportional to
the difference between the flux emitted when X-rays illuminate the disc and the intrinsic flux of the disc (see Eq.\,1 in K23). Consequently, we expect the amplitude of the transfer function to increase when a larger fraction of the observed flux is due to the thermalisation of X-rays in the disc. This could happen when the flux due to X-ray absorption increases or the intrinsic emission of the NT disc decreases.
Furthermore, we recall that the response function is normalised to the flash's total X-ray luminosity, which means that its amplitude may drop with increasing X-ray luminosity. 

\subsubsection{Dependence on wavelength}
The top panel of Fig. \ref{fig:gamma2_wave} shows \gammaff\ at three different wavebands, $\lambda=1300, 3000, 5000$\AA. The amplitude of the transfer function decreases with increasing wavelength. This is because the amplitude of the response function also decreases with increasing wavelength, due to the fact that the total difference between the flux of the illuminated disc and the intrinsic flux of the disc decreases at longer wavelengths. For our fiducial parameters, this difference peaks at around 1100\AA; that is, at wavelengths smaller than $1300$\AA, while it decreases at longer wavelengths. 
In general, the UV/optical variability amplitude in AGNs should be maximum at (approximately) wavelengths where the X-ray illuminated disc spectrum peaks, and then it should decrease both at shorter and longer wavelengths. We stress that this should be the case as long as the variability amplitude is not normalised to the mean flux. 

When we normalise the disc transfer function to the mean disc flux in each wavelength, the difference between the amplitude of the transfer function in various wavelengths is significantly reduced, as is seen in the bottom panel of Fig.\,\ref{fig:gamma2_wave}. This is because we divide the transfer function by a quantity (the square of the disc flux), which also becomes smaller as we observe at wavelengths further away from the disc peak emission. In the case B corona (dashed lines), the decrease in the amplitude of the transfer function is almost equal to the decrease of the disc emission, as both depend in the same way on the extra X-ray luminosity that shines the disc.


\subsubsection{Dependence on \texorpdfstring{\ltransf}{ltransf}}

The top left panel of Fig.\,\ref{fig:gamma2_params} shows \gammaff\ as a function of frequency for different values of \ltransf, at $\lambda=3000$\AA. 
In general, the amplitude of the transfer function is not strongly dependent on \ltransf. For example, in the case of the externally powered corona, the difference in amplitude between the \ltransf=-0.1 and \ltransf=-0.9 transfer functions (dashed lines in the upper left panel in Fig.\,\ref{fig:gamma2_params}) is less than $\sim $20\%. This is because the response function is normalised to the luminosity of the X-rays, which increases proportionally to \ltransf. This is similar to the case A corona, where we see that the amplitude of the transfer function increases by a factor of $\sim 2$, when \ltransf\ increases by almost an order of magnitude (i.e. from 0.1 to 0.9). 

We note that when \ltransf\ is negative (i.e. the corona is powered externally), the amplitude of \gammaff\ slightly decreases with increasing X-ray luminosity. This happens because the 
additional flux resulting from the reprocessed emission increases at a slower pace than $L_{X}$, as is explained in P22. For positive \ltransf, the amplitude of the disc transfer function increases with increasing \ltransf, which means that the flux of the reprocessed emission increases faster than $L_{X}$. This happens because in this case all the accretion power produced in radii smaller than \rtransf\ is given to the corona. As a result, the total disc emission below \rtransf\ is solely due to the reprocessing of X-rays, which causes peaks to appear in the disc response at times shorter than the time it takes for the X-ray flux to illuminate the disc up to \rtransf\, (see Fig.\,1 in K23). Their amplitude (and width) increases with increasing \ltransf, leading to an increase in the \gammaff\ amplitude. 

\subsubsection{Dependence on BH spin}
The upper middle panel of Fig.\,\ref{fig:gamma2_params} shows \gammaff\ for different values of \spin. In general, as previously with \ltransf, the amplitude of the transfer function does not depend strongly on BH spin. For example, the case B transfer functions (dashed lines) vary by a factor of $\sim 1.8$ between spin of 0 and 0.998, while this factor reduces to 1.2 in the case A corona (solid lines). 

We note that, in the case B corona, \gammaff\ increases with BH spin. 
Faster spins correspond to higher radiative efficiency, resulting in a lower accretion rate in physical units. Consequently, the temperature of the accretion disc drops, leading to a higher fraction of the observed emission being from X-ray reverberation. 

\subsubsection{Dependence on \texorpdfstring{\fcol}{fcol}}

The upper right panel of Fig.\,\ref{fig:gamma2_params} shows the disc transfer function for various \fcol. The colour correction factor significantly affects the amplitude of the transfer functions. 
The amplitude of \gammaff\ decreases with increasing \fcol, regardless of how the corona is powered. This is because the amplitude of \psif\ decreases with increasing \fcol\ (see Fig.\,1 in K23), since the temperature of each disc element increases with an increasing colour correction factor. Consequently, the intrinsic flux of the disc increases, leading to a smaller amplitude for \psif\ and the transfer function.

\begin{table*}
      \caption[]{Parameter values for which we computed the model power spectra.}
         \label{tab:parameters}    
\[\arraycolsep=6pt\def\arraystretch{1.2}
         \begin{array}{lll}
            \hline
            \hline
            \noalign{\smallskip}
            \textrm{Parameter} &  & \textrm{Values} \\
            \noalign{\smallskip}
            \hline
            \noalign{\smallskip}
            \mathrm{Wavelength} &
            \lambda_{\rm rest} (\text{\AA}) & 1300, 1800, 2300, 3000, 3400, 4000 \\ 
            
            \textrm{BH mass, accretion rate, redshift} & (\mathrm{log M}_{\rm BH},\dot{m}_{\rm Edd},\overline{z}) &
            (8.3, 0.1, 0.72), (8.3, 0.2, 1.04), (8.3, 0.4, 1.43) \\ 
             &  & (8.9, 0.04, 1.03), (8.9, 0.1, 1.44), (8.9, 0.25, 1.77) \\
             & & (9.5, 0.025, 1.44), (9.5, 0.05, 1.88), (9.5, 0.1, 1.82) \\
            
                 \textrm{X-ray luminosity} &
                  L_{\rm transf}/L_{\rm disc}  & \\
                  \textrm{(when corona is powered by the accretion process)} & &
                  0.1, 0.3, 0.5, 0.7, 0.9 \\
                 \textrm{(when corona is powered an external source)} &  & -1.9, -1.7, ..., -0.3, -0.1 \\
                 
                \textrm{Corona height}  &
                 h (R_{\rm g}) & 2.5, 5, 10, 20, ..., 90 \\
                 
               \textrm{Colour-correction factor} &
                f_{\rm col} & 1, 1.7, 2.5 \\
                
                \textrm{BH spin}  &
                \alpha^{*} & 0, 0.7, 0.998\\
            \noalign{\smallskip}
            \hline
         \end{array}
\]
\end{table*}

\subsubsection{The amplitude of the normalised disc transfer function}
\label{sec:gammanorm-amp}


The bottom panels in Fig.\,\ref{fig:gamma2_params} show transfer functions normalised to the square of the disc flux at the same waveband (these are the functions in which we are interested, since we aim to fit PSDs that are normalised to the mean flux squared). The left and middle panels show that the dependence of \gammanorm\ on \ltransf\ and \spin\ is stronger than the one on \gammaff. This is because the disc flux depends on these parameters more strongly than the reprocessed flux. In the case B corona, 
the disc flux increases with increasing \ltransf, and as a result, the amplitude of the normalised transfer function decreases with increasing \ltransf. In the case A corona, the disc flux is approximately the same for \ltransf<0.5 (at $\lambda=3000$\AA) but decreases significantly for  \ltransf=0.9, leading to a higher amplitude for \gammanorm. Similarly, the disc flux at $\lambda=3000$\AA\ decreases with increasing BH spin, and as a result the amplitude of \gammanorm\ increases.

Finally, the bottom right panel of Fig.\ref{fig:gamma2_params} shows that 
the difference between the amplitude of \gammanorm\ for various \fcol\ is smaller than the difference between the \gammaff\ amplitudes. This happens because the SED of the disc is shifted to higher energies with increasing \fcol\ values, leading to a decrease in the disc flux at $\lambda = 3000$\AA\ (for the fiducial values used here). As both the disc transfer function and the disc flux decrease with increasing \fcol\,, the overall dependence of \gammanorm\ on \fcol\ becomes weaker (and even non-existent in the case B corona).

\subsection{The break frequency of the disc transfer function}

The break frequency of the transfer function depends on the width of the response function. As \psif\ becomes wider, \gammaff\ becomes narrower; that is, its break frequency moves to lower frequencies. The width of the response function increases as the size of the disc region that contributes to a particular waveband increases. As a result, the break frequency decreases with increasing wavelength (see Fig.\,\ref{fig:gamma2_wave}), since the emission at larger wavelengths originates from a more extended disc region. 
In general, the size of the emitting region increases with increasing temperature. Consequently, changes in the system parameters that increase the disc temperature cause the response width to be larger.

The left panels of Fig.\,\ref{fig:gamma2_params} show that the break frequency of \gammaff\ slightly decreases with increasing \ltransf\. 
This is because the disc temperature increases as a result of increased X-ray heating. 
The middle panels of Fig.\,\ref{fig:gamma2_params} show that the break frequency of \gammaff\  slightly increases with increasing \spin. This is because higher spin values correspond to a lower accretion rate in physical units, which results in a lower disc temperature and consequently a narrower response function. We note, however, that \ltransf\ and \spin\ have a minimal effect on the break frequency.
The right panels of Fig.\,\ref{fig:gamma2_params} show that the break frequency of \gammaff\ decreases as \fcol\ increases. This occurs because for larger \fcol\ the disc temperature increases, causing the flux at a given waveband to be emitted from a larger disc region. As a result, the \psif\ broadens, leading to a narrower \gammaff.

\section{The observed UV/optical power spectrum of quasars}
\label{sec:obs}

\citet{Petrecca24} studied the ensemble PSD function of SDSS quasars in the UV/optical bands. They studied 8042 SDSS sources in the Stripe-82 region that cover a broad range of BH masses ($\sim 10^{7.5} - 10 ^{10}$ M$_{\odot}$), luminosities ($L_{\rm bol}\sim 10^{45} - 10 ^{47}$erg/s), and redshifts ($0.5 \leq z \leq 2.5$). They computed the rest-frame periodogram of each source at $\nu_{\rm obs}=4.6\times 10^{-4}, 9.1\times 10^{-4}$ and $1.4\times 10^{-3}$ days$^{-1}$
(i.e. on timescales of $t_{\rm obs}=2190, 1095, 730$ days) using binned light curves with six equidistant points. Then, they grouped the quasars into narrow $[M_{\rm BH}, L_{\rm bol}]$ bins and computed the mean power spectrum of all quasars in each bin. This was done for seven rest-frame wavebands with centroids of $\lambda_{\rm rest} = 1300, 1800, 2300, 3000, 3400, 4000, 5000 $\AA\ and widths $\Delta \lambda$, so that $\Delta \lambda / \lambda_{\rm rest} = 0.2$. Each $[M_{\rm BH}, L_{\rm bol}, \lambda_{\rm rest}]$ bin contains quasars at all redshifts, since P24 did not find strong evidence of a PSD dependence on redshift.

\cite{Petrecca24} showed that the (rest-frame) power spectra of quasars can be fitted well by a power-law model of the form
\begin{equation}
    \textrm{log}[\textrm{PSD}_{\lambda, \rm obs}(\nu)] = \textrm{log}[\textrm{PSD}_{\lambda, \textrm{amp}}] + \textrm{PSD}_{\lambda, \textrm{slope}}[\textrm{log}(\nu)+2.6] \;,
    \label{eq:psd_fit}
\end{equation}
where $\textrm{PSD}_{\lambda,\textrm{amp}}$ is the power spectrum amplitude at $\nu = 10^{-2.6}$ day$^{-1}$, and $\textrm{PSD}_{\lambda,\textrm{slope}}$ the slope of the power spectrum. They found that $\rm log[PSD_{\lambda,amp}]$ and $\rm PSD_{\lambda,slope}$ depend linearly on $\rm log(M_{\rm BH})$ and $\rm log(\lambda_{\rm Edd})$ for each of the rest-frame wavelengths studied (see Eqs.(9) and (10) in P24), where $\lambda_{\rm Edd} = L_{\rm bol}/L_{\rm Edd}$ is the Eddington ratio, and $L_{\rm Edd}$ the Eddington luminosity. 

Our objective is to investigate whether X-ray disc reverberation can explain the P24 results. To this end, we considered three BH masses and three accretion rates \medd\
\footnote{P24 used the observed Eddington ratio, $\lambda_{\rm Edd}$ in their study. However, the K23 code that computes the response functions takes as input the accretion rate \medd. We assume that the Eddington ratio is a good approximation of the accretion rate, i.e. $\lambda_{\rm Edd} \sim \dot{m}_{\rm Edd}$.} for each BH mass, which are listed in Table\,\ref{tab:parameters}. We chose the smallest, the mean, and the largest BH mass in Fig.10 of P24, together with the appropriate range of accretion rates for each BH mass, to cover the whole parameter space studied by P24.

\begin{figure*}
   \centering
    \includegraphics[width=18cm, height=11cm]{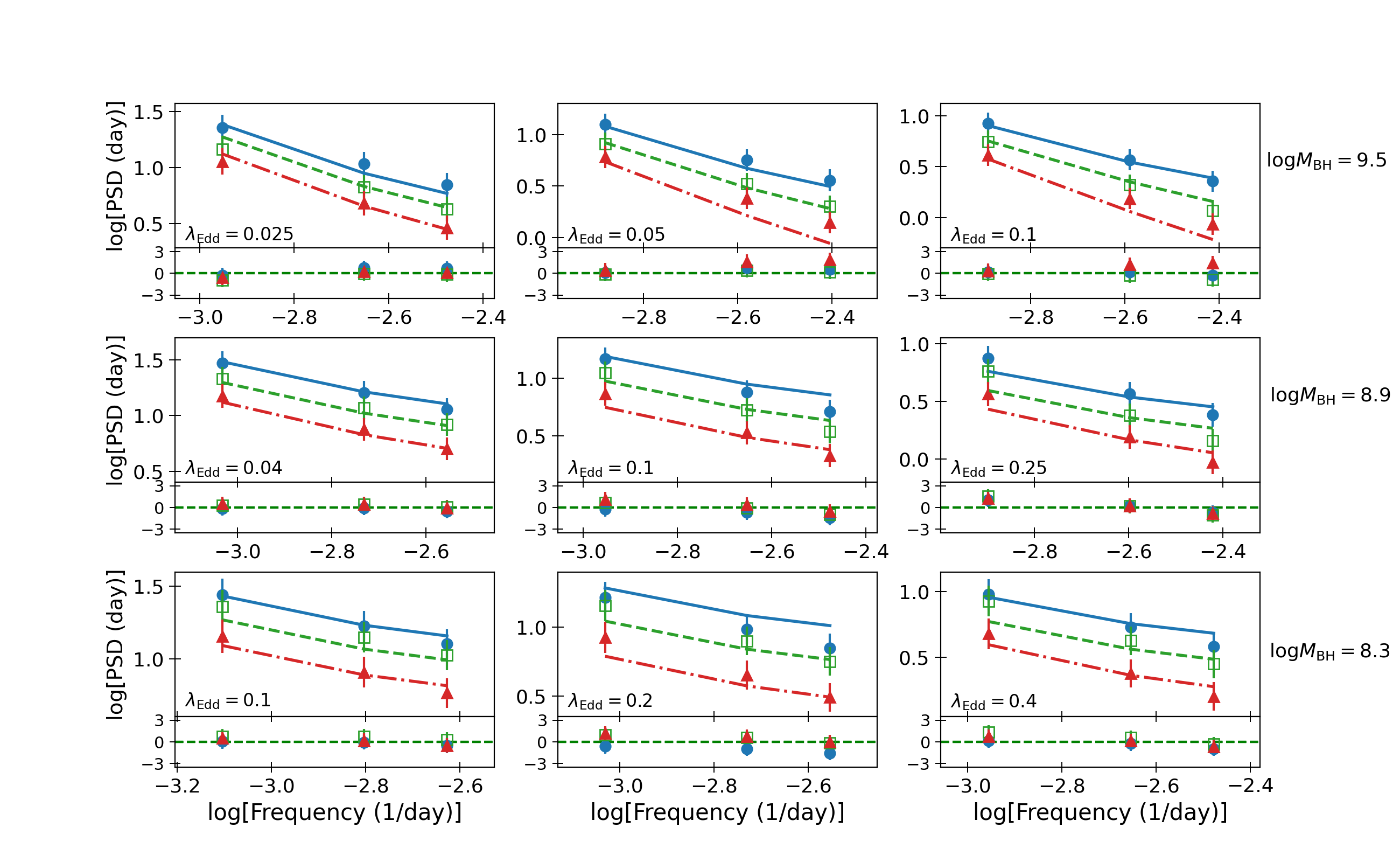}
      \caption{Blue circles, green squares, and red triangles indicate the observed PSDs at $\lambda$=1300\AA, 2300\AA, 4000\AA, respectively, for all combinations of BH mass and accretion rate we considered. Solid blue lines, dashed green lines, and dash-dotted red lines indicate the best-fit models for \spin=0, in the case in which the corona is powered by the accretion process (note that, for clarity reasons, the ranges of the axes are different in each plot). Error bars indicate the combined error of the data and model. The best-fit residuals (i.e. (data-model)/error) are shown in the bottom panel of each plot (see the top panel of Fig.\,\ref{fig:residuals} for the residuals at all wavelengths).}
         \label{fig:psd_fits}
   \end{figure*}

We computed the observed PSD for each combination of (\mbh, \medd) at the respective rest-frame frequencies of $\nu_{rf} =(1+\overline{z}) \nu_{\rm obs}$, where $\overline{z}$ is the mean redshift of each BH mass and accretion rate bin, as is listed in Table \ref{tab:parameters} (the mean redshift was calculated using the redshift of the individual sources in each bin). 
Blue circles, green squares, and red triangles in Fig.\,\ref{fig:psd_fits} indicate the observed power spectra as a function of frequency for all combinations of BH mass and accretion rates, at three different wavelengths. 


\section{The model power spectrum}
\label{sec:psd}

To fit the P24 results, we computed the model power spectra using Eq.\,(\ref{eq:psds}). To do this, we needed the transfer function \gammaff, which we calculated in the manner explained in Sect.\,\ref{sec:transfer}, and the X-ray power spectrum, which we computed in the manner explained below.

\subsection{The model X-ray power spectrum}
\label{subsec:psd_X}

The X-ray power spectrum of the SDSS quasars studied by P24 
is not known. Therefore, we adopt a PSD shape based on previous studies of nearby Seyferts \cite[e.g.][]{Uttley02,Markowitz03,McHardy04,Gonzalez-Martin2012}. Specifically, we assume that the X-ray PSD follows a bending power-law model, with a low-frequency slope of $-1$ and a high-frequency slope of $-2$, i.e.
\begin{equation}
    \textrm{PSD}_{\rm X}(\nu) = \frac{A}{\nu} \cdot \frac{1}{1+ \nu / \nu_{\rm b, X}} \;,
\label{eq:PSD_X}
\end{equation}
where $A$ is the normalisation and $\nu_{\rm b, X}$ the bending frequency. The PSD in Eq.\,(\ref{eq:PSD_X}) has units of per hertz. The normalisation, $A$, has no units and is assumed to vary with the accretion rate as 
\begin{equation}
    A = 6 \times 10^{-3} \dot{m}^{-0.8}_{\rm Edd}\;,
\label{eq:PSD_X_amp}
\end{equation}
following \cite{Ponti_2012} and \cite{Paolillo17}.  
To compute the bend frequency, $\nu_{\rm b, X}$, we assume the \cite{McHardy06} relation:
\begin{equation}
    \textrm{log}T_{\rm b,X} = a \times \textrm{log}(M_{\rm BH,6}) - b \times \textrm{log}(L_{\rm bol,44}) + c\;,
\label{eq:logTb}
\end{equation}
where $a = 2.1$, $b = 0.98$, $c = -2.32 $, $T_{\rm b, X}=1/\nu_{\rm b,X}$ (in days), M$_{\rm BH,6}$ is the BH mass in units of $10^{6}$M$_\odot$, and $L_{\rm bol,44}$ is the bolometric luminosity in units of $10^{44}$erg/s. The bolometric luminosity is equal to $L_{\rm bol} = \dot{m}_{\rm Edd} \cdot L_{\rm Edd}$.

The disc responses that we use are normalised to the total X-ray luminosity in Eddington units. This means that the quantity $F_{X}$ in Eq.\,(\ref{eq:discflux}) is the total X-ray luminosity measured in units of the Eddington luminosity, and  $\rm PSD_{X}$ in Eq.\,(\ref{eq:psds}) is the power spectrum of this $F_{X}$.
However, Eq.\,(\ref{eq:PSD_X}) defines the power spectrum of a light curve normalised to its mean.
For this reason, we need to multiply the PSD model defined by Eq.\,(\ref{eq:PSD_X}) by the X-ray luminosity in Eddington units, $L_{X, \rm Edd}$, squared. This luminosity was calculated using the {\tt KYNXiltr} code for any combination of the model parameters that we considered.

\subsection{The model UV/optical power spectrum}

\begin{figure}
   \centering
    \includegraphics[width=9cm, height=7cm]{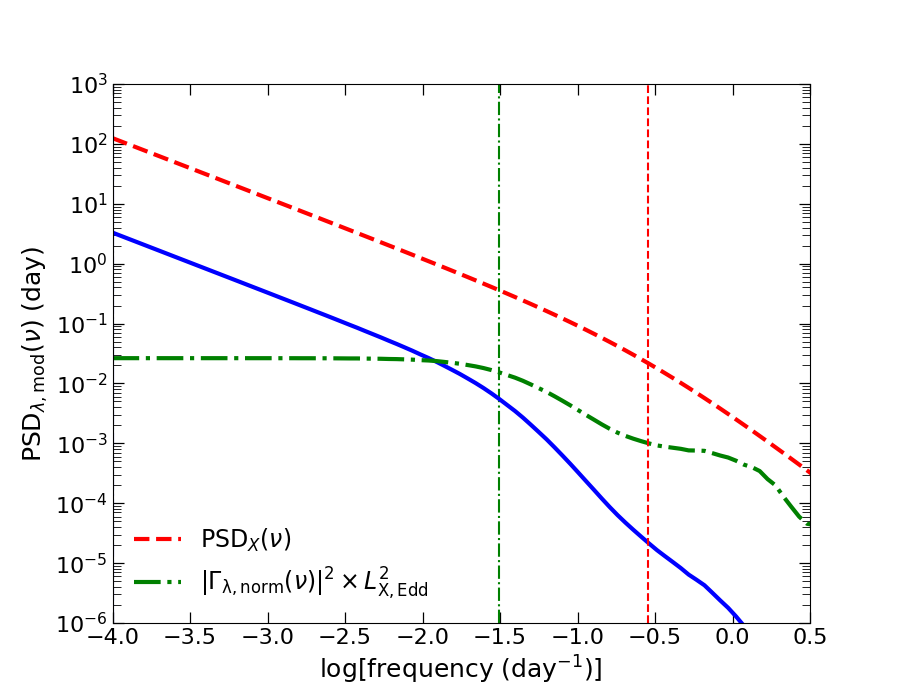}
      \caption{Model power spectrum at $\lambda=3000$\AA\ (solid blue line) for the parameters: $M_{\rm BH}=2 \times 10^8 M_{\odot}$, \medd\ = 0.4, $h=20$, $\Gamma=2$, $\theta = 30^o$, $\alpha^* = 0$, $R_{\rm out}=5000 R_{\rm g}$, $E_{\rm cut} = 150$keV, $L_{\rm transf}/L_{\rm disc}=0.5$, and $f_{\rm col}=1.7$. The dashed red line and the dash-dotted green lines indicate the corresponding X-ray power spectrum, $\textrm{PSD}_{\rm X}(\nu)$, and the normalised transfer function (multiplied by $L^2_{X, \rm Edd}$ so that the product of the red and green line is equal to the blue line as Eq.(\ref{eq:psd_mod}) shows), respectively. The dashed red and vertical dash-dotted green lines show the position of the X-ray PSD break frequency and the transfer function break frequency, respectively.}
         \label{fig:psd_example}
   \end{figure}

We calculated the UV/optical PSD using Eq.\,(\ref{eq:psds}), with the transfer function and the X-ray PSD given by Eqs. (\ref{eq:fourier_discrete}) and (\ref{eq:PSD_X}), respectively. This gives the PSD in units of day $\times$ (erg/s/cm$^2$/\AA)$^2$. However, P24 computed the PSD by normalising the light curves to their mean, so their PSDs are in units of (day). Therefore, it is necessary to also normalise the model PSDs with the square of the mean flux expected in each spectral band. To do this, we used \gammanorm\ (defined in Eq.\, (\ref{eq:gnorm})) to calculate the model PSDs. 
Consequently, the model power spectrum in the UV/optical bands is given by
\begin{equation}
    \textrm{PSD}_{\rm \lambda, mod}(\nu) = |\Gamma_{\rm \lambda, norm}(\nu)|^{2} \cdot \textrm{PSD}_{\rm X}(\nu) \cdot  L_{\rm X,Edd}^2\; .
\label{eq:psd_mod}
\end{equation}
Figure\,\ref{fig:psd_example} shows an example model power spectrum 
at $\lambda = 3000$\AA\ (solid blue line). We also plot the respective X-ray power spectrum, \psdX, with the dashed red line, and the transfer function normalised to the flux of the disc \gammanorm\ (multiplied by $L^2_{X, \rm Edd}$), with the dash-dotted green line. The vertical dashed red line shows the position of the X-ray power spectrum break frequency, given by Eq.\,(\ref{eq:logTb}), and the vertical dash-dotted green line shows the position of the break of \gammaff. Given the two different break frequencies, the model UV/optical PSD shows two breaks, one due to the X-ray PSD and the other due to the disc transfer function. The location of the two breaks in the UV/optical PSD is not fixed and depends on the BH mass and the accretion rate of the source. The break due to the X-ray PSD appears after the \gammaff\ break in Fig.\,\ref{fig:psd_example}, but this will not be the case at all times. 
As the accretion rate decreases, the break of \psdX\ moves to lower frequencies, while the break of \gammaff\ moves to higher frequencies (see Fig.10 in P22). At frequencies higher than both breaks, the slope of the power spectrum is steeper than -2. 


\subsection{Aliasing corrections}
\label{subsec:alias}

The observed power spectra are prone to aliasing. We need to take this
into account before comparing our model power spectra with the observed ones. Aliasing occurs when a continuous random process is sampled at fixed time intervals; for example, $\Delta t$. If the sampling rate is insufficient relative to the highest-frequency components of the signal, higher-frequency signals can become indistinguishable from lower-frequency signals in the sampled data. This results in the overlap of higher-frequency power into the lower-frequency range, leading to an observed power spectrum that is flatter than the intrinsic power spectrum. For this reason, we need to correct the model PSD so that they include the extra power due to aliasing as well. To do this, we followed \cite{Priestley1981}, who showed that when a continuous process is sampled every $\Delta t$ the resulting power spectrum, $\textrm{PSD}^{\rm alias}_{\rm \lambda, \rm mod}(\nu)$, is related to the intrinsic power spectrum of the process, $\textrm{PSD}_{\rm \lambda, \rm mod}(\nu)$, as follows:
\begin{equation}
    \textrm{PSD}^{\rm alias}_{\rm \lambda, \rm mod}(\nu) = \sum_{k=0}^{\infty} \textrm{PSD}_{\rm \lambda, \rm mod}(\nu \pm k/\Delta t) \;.
    \label{eq:aliasing}
\end{equation}
The rest-frame sampling rate of the power spectra of P24 is different for each source, since they studied sources at a wide range of redshifts. For this reason, we consider a rest-frame sampling rate of $\Delta t = 365/(1+\overline{z})$, where the values of $\overline{z}$ are listed in Table\,\ref{tab:parameters} for each of the ($M_{\rm BH}$, \medd) cases we consider. Furthermore, P24 binned light curve points in bins of size $\sim 3$ months, which corresponds to a rest-frame binning of $\sim 90/(1+\overline{z})$ days. This procedure reduces the amount of power that is aliased back to the sampled frequencies from timescales shorter than $90/(1+\overline{z})$ days \cite[e.g.][]{Van_der_klis88}. For this reason, we corrected our model power spectra by computing the sum of Eq.\,(\ref{eq:aliasing}) only up to $k$ which satisfies $|\nu \pm k/\Delta t |\leq (1+\overline{z})/90$ day$^{-1}$. All model power spectra computed below are corrected for aliasing effects.

\section{Fitting the observed PSD}
\label{sec:fitting}


 \begin{figure*}
   \centering
    \includegraphics[width=18cm, height=5.2cm]{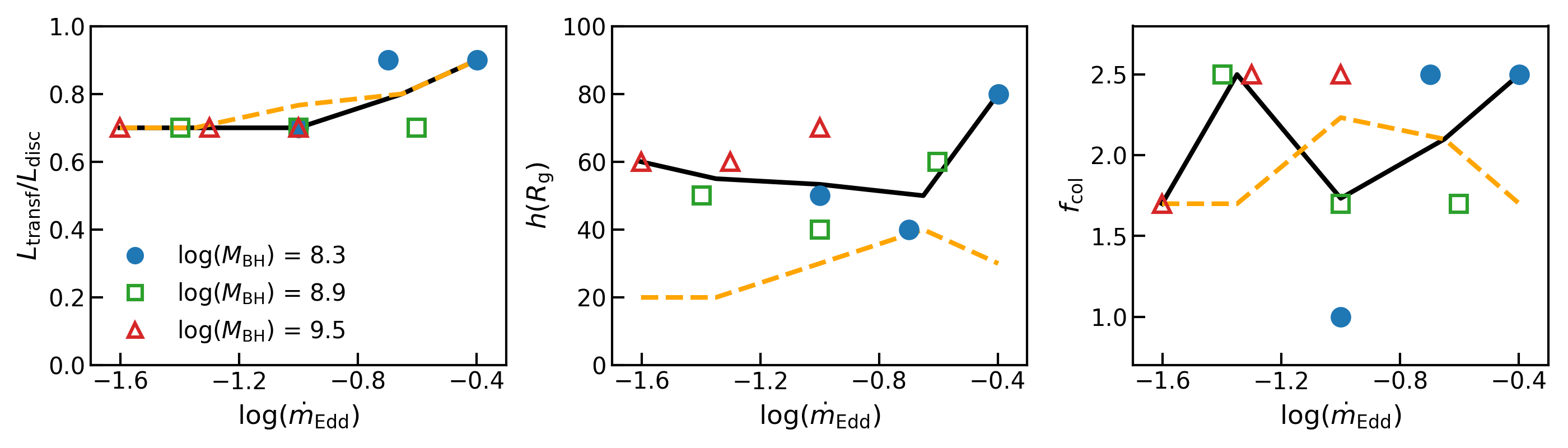}
      \caption{Best-fit results in the case A corona. The \spin=0 best-fitting \ltransf\ (left panel), $h$ (middle panel), and $f_{\rm col}$ (right panel) are plotted as a function of \medd\ for log$M_{\rm BH}=8.3$ (blue circle points), log$M_{\rm BH}=8.9$ (green square points), and log$M_{\rm BH}=9.5$ (red triangle points).  
Black lines show how the best-fit parameters change with the accretion rate (we bin points that have similar accretion rates) for the \spin=0 models. The dashed orange lines correspond to the \spin=0.7 models. }
    \label{fig:best_fit_params}
\end{figure*}


We calculated the model power spectra for the $\lambda_{\rm rest}$, $M_{\rm BH}$, \medd\ values listed in Table \ref{tab:parameters}. They are the same values we used to calculate the observed PSDs as well, except for $\lambda_{\rm rest}=5000$\AA. We chose not to fit the PSD in this waveband because the number of quasars with observations in this band is smaller than the rest (at least five times smaller). Consequently, the PSD parameters at this wavelength are determined in a smaller BH mass and accretion rate range than the rest of the wavelengths. We considered the X-ray corona heights, X-ray luminosities, colour-correction factors, and BH spins that are listed in Table \ref{tab:parameters}. The rest of the model parameters are set to the fiducial values (i.e. $\Gamma=2, \theta =30^o, R_{\rm out}=5000R_{\rm g}, E_{\rm cut}=150$keV). We created model power spectra for all combinations of the above parameters, and thus formed a structured grid to fit the data. The step size of the parameters was chosen to ensure sufficient coverage of the parameter space while maintaining a manageable runtime.

We fitted the PSDs separately for the case A and case B scenarios. 
In each scenario, we assume the same BH spin for all the nine M$_{\rm BH}$ and $\dot{m}_{\rm Edd}$ bins that we considered. For each pair of (M$_{\rm BH}$, $\dot{m}_{\rm Edd})$, we fitted the observed power spectra at all wavelengths, PSD$_{\rm \lambda,obs}$, letting \ltransf, $h$, and $f_{\rm col}$ be free parameters, since they may depend on the BH mass and the accretion rate. 
To determine the best-fit parameters in each (M$_{\rm BH}$, $\dot{m}_{\rm Edd})$ bin, we minimised the traditional $\chi^2$ statistic, defined as follows:
\begin{equation}
    \chi^{2} _{({\rm M}_{\rm BH}, \dot{m}_{\rm Edd})} = \sum_{i=1}^{6} \sum_{j=1}^{3}\frac{\{\rm log[PSD_{\lambda_{i}, \rm obs}(\nu_{j})] - \rm log[PSD_{\lambda_{i}, \rm mod}(\nu_{j})]\}^{2}}{\sigma^{2}_{\lambda, \rm obs}(\nu_{j}) + \sigma^{2}_{\lambda, \rm mod}(\nu_{j})}\;,
    \label{eq:chi2}
\end{equation}
\noindent where index $i$ runs on all wavelengths and $\nu_{j}$ are the three rest-frame frequencies ($\nu_{\rm rf} = (1+\overline{z})\nu_{\rm obs}$), as is explained in Sect.\ref{sec:obs}. 

The term $\sigma^2_{\rm \lambda, obs}(\nu_j)$ indicates the error of the observed PSDs. These errors are due to the error of the best-fit coefficients that determine PSD$_{\rm \lambda,amp}$ and PSD$_{\rm \lambda,slope}$ in Eq. (\ref{eq:psd_fit}), as is reported by P24 in their Table 1, and we computed them through standard error propagation. We also averaged the observed errors over the various wavelengths to ensure that the fit of the model would not be determined by the best-fit results of \cite{Petrecca24} for a specific wavelength. The term $\sigma^{2}_{\lambda, \rm mod}(\nu_{j})$ indicates the error (uncertainty) of the model PSDs. This arises (mainly) from the fact that we cannot determine the X-ray PSDs of the quasars in the sample. We assumed a low- and high-frequency PSD slope of $-1$ and $-2$ for all quasars, which is probably an oversimplification. Even if this were the case, we use relations to compute the PSD break frequency and the PSD amplitude, which have relatively large uncertainties. Furthermore, our aliasing prescription is simplified and should add some extra uncertainty in the model predictions, as the bin size of the light curves that P24 used to compute the PSDs was not the same even for all the points within the same light curve. For all these reasons, we assume an error of 0.1 dex due to the uncertainties of the power spectrum model\footnote{We got this estimate for the error due to the X-ray PSD model uncertainties by computing the range in the resulting UV/optical model PSDs when we considered the 1$\sigma$ error of the parameters that determine the dependence of the PSD break frequency and PSD amplitude on BH mass and/or accretion rate.}.
 
The final, overall best-fit in the case A or case B scenarios is the one that minimises the global chi-square, $\chi_{g}^{2}$, defined as follows:
\begin{equation}
    \chi^{2}_g(\alpha^*_k,L_{\rm transf}/L_{\rm disc}^{(ij)},h^{(ij)},f^{(ij)}_{\rm col})=\sum_{i=1}^{3} \sum_{j=1}^{3} \chi^2_{M_{\rm BH}(i), \dot{m}_{\rm Edd}(j)},
    \label{eq:chi2-global}
\end{equation}
\noindent where $k=1,2,3$ are the three BH spins we consider in each case, while $L_{\rm transf}/L_{\rm disc}^{(ij)},h^{(ij)},$ and $f^{(ij)}_{\rm col}$ are the best-fit parameters for each of the nine pairs (M$_{\rm BH}$, $\dot{m}_{\rm Edd})$ we considered. 

The observed power spectra were calculated at three frequencies for each of the six wavelengths, so for each ($M_{\rm BH}$, $\dot{m}_{\rm Edd})$ there are 18 points. 
Therefore, the total number of degrees of freedom for the global fit is $9 \; (\textrm{pairs of } M_{\rm BH}, \dot{m}_{\rm Edd})  \times (18-3 \textrm{ free parameters})=135$. 

\section{The best-fit results}
\label{sec:best_fit}


\subsection{Case A: Accretion-powered X-ray corona} 

In the case in which the X-ray source is powered by the accretion process (\ltransf>0), the best-fit results are: $\chi^2_{g, \rm min}=110$ for \spin=0,  
113 for \spin=0.7, 
and 151 for \spin = 0.998 (for 135 degrees of freedom). 
The \spin = 0 and \spin = 0.7 models fit the PSDs equally well. On the other hand, although the \spin=0.998 model fits are statistically acceptable (the null hypothesis probability is $\sim 0.16$) they are significantly worse ($\Delta \chi^2>38$, for both the \spin=0 and \spin=0.7 model fits)\footnote{The evidence ratio, $\epsilon=e^{-\Delta\chi^2/2}$, is a measure of the relative likelihood of two models that fit the data with the same degrees of freedom \citep[see e.g.][]{Emmanoulopoulos16}. The fact that $\Delta\chi^2>38$ suggests that the spin 0 and 0.7 models are significantly more likely than the maximum spin model.}. This result suggests that, within the X-ray reverberation model, maximally spinning BHs are probably rare in the sample of quasars studied by P24.

Figure\,\ref{fig:residuals} shows the best-fit residuals 
for all wavelengths and for all ($M_{\rm BH}$,\medd) cases, for the three BH spin values. 
Figure \ref{fig:psd_fits} shows the observed power spectra at $\lambda=1300$\AA, 2300\AA, and 4000\AA, together with the best-fit models, and the respective residuals for the \spin=0 case. This figure demonstrates the overall goodness of fit to the actual PSDs (we get similar plots for the spin 0.7 case).

The points in Fig.\,\ref{fig:best_fit_params} show the best-fit values of the parameters \ltransf, $h$, and $f_{\rm col}$  plotted as a function of \medd\ (left, middle, and right panel, respectively) for the case \spin=0. The black line shows how the best-fit parameters change with the accretion rate when we bin points that have similar accretion rates. The figures show that \ltransf$\sim 0.7-0.8$, for all BH masses and accretion rates.  
The best-fit height of the X-ray corona appears to be $\sim 60 R_{\rm g}$. Both \ltransf\ and the corona height may increase for the highest accretion rate, but this trend is suggestive at best. 
We do not observe a clear trend of the best-fit \fcol\ with the accretion rate. Our results indicate a mean \fcol\ of $\sim 2.$. This is slightly higher than recent predictions \citep[see e.g.][]{Davis_2019}, which suggest a colour correction factor of $\sim 1.6-1.8$, for the BH mass and accretion rate of the AGN in the sample of P24. 

The dashed orange lines in Fig.\ref{fig:best_fit_params} show the same parameters for the \spin = 0.7 best-fit models. 
They are similar to the best-fit parameters when \spin=0, except that the best-fit height of the corona is smaller. We find that $h$ increases from $\sim 20$ to $40 R_g$ with an increasing accretion rate. This is expected since the UV/optical variability amplitude in the case of X-ray reverberation increases with increasing \spin (see the bottom middle panel of Fig.\,\ref{fig:gamma2_params}). Consequently, to match the observed variability, the model requires a smaller height for the X-ray source, since the variability amplitude decreases with decreasing corona height (as is shown in Fig. 8 of P22).


\subsection{Case B: Externally powered X-ray corona}

The case B model PSDs cannot fit the observed power spectra. This is because the amplitude of the transfer function normalised by the accretion disc flux (squared) in case B coronae is significantly smaller than in case A, as is shown, for example, in the bottom left panel of Fig.\,\ref{fig:gamma2_params}. Since the best-fit value of \ltransf\ in the case A corona turns out to be $\sim 0.8$, the left panel in Fig.\,\ref{fig:gamma2_params} shows that the amplitude of \gammanorm\ in case B will be more than 10 times lower even if |\ltransf|=0.1 (and significantly smaller if |\ltransf| is larger). Our results show that the large amplitude of the observed PSDs, when normalised to the disc flux (squared), cannot be explained by an externally powered corona, as the flux of the accretion disc in this case is far too large for the predicted variability.

\section{Summary and discussion}
\label{sec:discuss}

In this work, we have modelled the UV/optical power spectra of quasars within the X-ray reverberation scenario. To do so, we used the code of \cite{Kammoun23} to calculate the response functions of the accretion disc for various parameters of the model and then used these functions to calculate the transfer function of the disc. Our approach is similar to P22, but the \cite{Kammoun23} code allows us to explore a larger parameter space. We computed the disc transfer function in the case in which the X-ray corona is powered by the accretion process or by an external source of power, for any value of \fcol, the BH spin, and the outer radius of the disc. We studied in detail the dependence of the transfer function, both in physical units and when it is normalised to the disc flux, on the wavelength, \ltransf, BH spin, and \fcol, both in the case of the externally powered X-ray corona and the accretion-powered one.

We fitted the observed UV/optical power spectra of AGNs 
using the recent results of P24, who determined the PSD spectra of the quasars in the SDSS Stripe-82 field, from (rest frame) $1300$\AA\ to 5000\AA. In order to fit the data, we did the following: 

1) We considered nine combinations of BH mass and accretion rate, which are listed in Table \ref{tab:parameters}, and determined their observed PSD, following Eq.(\ref{eq:psd_fit}) and Eqs.(9) and (10) in P24. 

2) We assumed that the X-ray PSD of the quasars follows a bending power-law form, with a low- and high-frequency slope of $-1$ and $-2$, respectively, and that the bending frequency depends on both the BH mass and the accretion rate, determined by \cite{McHardy06}, and c) that the X-ray PSD amplitude is equal to $0.06 \times \dot{m}^{-0.8}_{\rm Edd}$ \cite[e.g.][]{Ponti_2012}.

3) We fitted the data for both an accretion-powered X-ray corona and an externally powered X-ray corona for a large number of model parameters (see Table \ref{tab:parameters}).

We found that X-ray reverberation can fit the observed power spectra well, but only when the corona is powered by the accretion power. The hypothesis of an externally powered corona is rejected by our results. This is an important result, as it shows that, under the hypothesis of X-ray reverberation, the mechanism that powers the X-ray corona in AGN should be associated with the accretion disc. 

The X-ray reverberation model can explain the P24 results in case A
for \spin=0 (and \spin=0.7), as long as \ltransf$\sim 0.8$, and the 
height of the X-ray corona is $\sim 60 R_g$ ($\sim 20-40 R_g$). 
Our results show that \fcol$\sim 2$, which is a value expected in objects with a high accretion rate \cite[see e.g.][]{Ross92}.

We note that the agreement between the X-ray reverberation model and the observed UV/optical PSDs of quasars is quite remarkable given the various simplifications of the model. First of all, the X-ray PSD of the quasars in the sample is unknown, and we had to assume a PSD model based on observations of nearby Seyferts. The X-ray PSD directly affects the model UV/optical PSD spectrum (see e.g. Eq.\,\ref{eq:psds}). It determines the location of the frequency breaks and the high-frequency slopes of the UV/optical power spectra.
It also affects our ability to model properly the amount of aliasing present in the observed PSDs. 
In addition, we fit the observed PSDs by keeping the BH spin fixed to the same value for all BH mass and accretion rate bins. Although we should not expect a relation between \spin\  and BH mass or $\dot{m}{\rm _{Edd}}$, we should expect a distribution of BH spins among the quasars in the P24 sample. As spins also affect the UV/optical PSD amplitude, we expect some of the scatter in the panels of Fig.\,\ref{fig:residuals} to be due to this simplified approach regarding the BH spin of the quasars in the P24 sample. 

In summary, our main result is that the X-ray reverberation model can fit well the power spectra of quasars in the Stripe-82 field in all wavelengths, from the far UV up to $4000$\AA, especially if we consider 
the various approximations we had to make (for the X-ray PSD, the BH spin being the same in all quasars etc). 
We note that X-ray reverberation of the accretion disc can also explain the apparent discrepancy between the observed half-light radii of X-ray illuminated discs and the size of the standard accretion disc models \citep{Papadakis22}, and the observed UV/optical time lags in a large sample of quasars \citep{Langis24}. These results support the hypothesis that the (majority of the) observed UV/optical variability in quasars is due to X-ray illumination of standard accretion discs.

An assumption of the model is that the X-ray-emitting corona is approximated as a point source located at a certain height along the BH's rotational axis (i.e. the lamp-post geometry). The energy spectrum emitted by a point source approximation (adopted by {\tt KYNSED}) is a reasonable representation of the energy spectrum from an extended 3D corona, as is shown by \citet[Section 5.3]{Dovciak22}, who compared the output of {\tt KYNSED} with that of the 3D Monte Carlo radiative transfer code {\tt MONK} \citep{Zhang19}. Furthermore, the effective size of the corona derived a posteriori by {\tt KYNSED} (by enforcing the conservation of the photon number during the Comptonisation process) is very similar to the true radius of a 3D spherical corona that emits the same energy spectrum as the one determined by {\tt MONK}. However, the X-ray reverberation output in the UV/optical bands depends not only on the energy spectrum but also on the illumination pattern of the disc (as a function of radius). \citet{Zhang24} have demonstrated that the illumination profiles of accretion discs irradiated by stationary spherical coronae remain largely insensitive to the coronal size at distances larger than $10-20R_g$ from the central BH. Therefore, our results are valid for any 3D coronae that have a rough spherically symmetric geometry.

We note that the significant X-ray polarisation detected by IXPE in NGC 4151 appears to favour flattened corona geometries, such as slab- or wedge-shaped configurations \citep{Gianolli23}. However, this is the case when one considers the expected polarisation only due to the inverse Compton process in the X-ray corona. However, it is important to note that disc reflection from a lamp-post geometry can also contribute to the observed polarisation signal. For example, \citet{Podgorny23} demonstrated that, in a lamppost set-up with unpolarised primary emission with a photon index of $\Gamma = 2$ and an inclination angle of $60^\circ$, the total degree of polarisation in the 2–8 keV band can reach values as high as $\sim 4 - 7\%$ (see their Table\;2). This is similar to what is detected in NGC 4151. Therefore, the geometry adopted in this work may still be consistent with the recent X-ray polarimetric observations of AGNs.

\subsection{The intrinsic variability of the accretion disc and other sources of UV/optical variability in quasars}

It is possible that the observed UV/optical variability of quasars can be caused by intrinsic variations of the accretion disc as well. For example, \cite{Lyubarskii97} studied the case of an accretion disc with the viscosity parameter fluctuating at different radii. If the resulting fluctuations of the accretion rate at different radii are not correlated, and of the order of the local viscous timescale, then the power spectrum of the UV/optical variations should have a power law-like shape, with a slope of $-1$. 

To investigate the potential contribution of such variations to the observed power spectra, we added to our model power spectrum a power-law component of the form PSD$_{\rm intr}(\nu)=A_{\lambda}\times (\nu / 10^{-2.6})^{-1}$, and we re-fitted the observed PSDs in the case of the corona powered by accretion with \spin = 0. We left the normalisation of the power law to be free with the wavelength. The best fits improve for all cases of the BH mass and accretion rate by approximately $\Delta \chi^2 = 6$. However, this improvement is not significant because it comes with the addition of six free parameters for each BH mass and accretion rate case. Furthermore, in all cases, the amplitude of the intrinsic disc variations is smaller than the one due to reverberation by at least a factor of 3. We conclude that X-ray reverberation can explain all the observed UV/optical variations in the Stripe-82 quasars in the frequency range spanned by our data.

This result may not be surprising given the long-term nature of the intrinsic disc variations. For example, the viscous timescale, $t_{\rm visc}$, of a thin accretion disc at about 10 Schwarzschild radii is $\sim 1400$ days for an AGN with \mbh=$2\times 10^8$M$_{\odot}$ \citep[assuming that the disc radius over height ratio is 0.1, and the viscosity parameter is 0.1; see Eq. 7 in][]{czerny06}. This is the smallest BH mass for the AGN we considered, so the viscous timescales should be longer in all of them.

The disc variability component in the UV/optical PSD will vanish at frequencies higher than $\nu_{\rm visc}=1/t_{\rm visc}$. Since log$(\nu_{\rm visc})\sim -3.2$ for $t_{\rm visc}=1400$ days and the lowest frequency sampled by P24 is $\sim -3.1$ (see Fig.\,\ref{fig:psd_fits}), then it will not be possible to detect the PSD intrinsic variability of the disc.  

Another component that can contribute to the observed variability is the Balmer continuum, which appears in the 2000-4000\AA\ range and is due to diffuse emission of gas in the broad line region. Recently, studies of the time lags between the UV/optical continuum variations showed indications that the variability of this component may be significant in this energy range. In fact, \cite{Netzer22} proposed that the variability of diffuse emission from radiation pressure-supported clouds in the broad line region with a covering factor of about 0.2 could explain entirely the observed continuum time lags in nearby Seyferts, if the ratio of the diffuse to the continuum emission at 3000\AA\ is $\sim 0.43$ (see their figure 2). This result would imply that the majority of the observed variations at wavelengths 2000-4000 \AA\ should be due to the variable Balmer emission component (and not to X-ray processing on the disc). However, the observed diffusion over continuum emission ratio in SDSS quasars at that wavelength is no more than 0.1-0.15 on average \citep[e.g.][]{Calderone17,Rakshit20}. Therefore, the variable emission from this component can only contribute a small part of the observed variations in the SDSS quasars that \cite{Petrecca24} studied. Furthermore, the fact that we do not detect a systematic increase over the best-fit models in the residual plots in Fig.\,\ref{fig:residuals} suggests that the contribution of this component in the observed PSDs may not be significant.

An additional potential source of UV/optical variability that could influence the measured power spectra in our sample is tidal disruption events (TDEs). These events typically evolve over timescales ranging from weeks to months for the rise and early decline phases, with some exhibiting extended tails lasting up to years \citep[see e.g.][]{vanVelzen20, Gezari21}. Given the timescale of these phenomena, there could be a power contribution to the frequency range probed in our analysis. However, our sample consists of quasars with BH masses exceeding $10^{8.3}M_\odot$. The maximum BH mass capable of tidally disrupting a solar-type star is approximately $10^8M_\odot$ for non-spinning BHs. Although this limit increases to nearly
$10^9M_\odot$ for maximally spinning BHs, the rate of TDEs declines significantly with increasing BH mass \citep[see e.g. Fig. 7 in][]{Mummery24}. Therefore, given the high BH masses in our sample, we consider the likelihood of contamination from TDEs to be negligible, and we do not expect them to affect our results.

\subsection{A universal shape for the power spectrum of quasars}
\label{sec:rescaled_psd}

As we already mentioned in the introduction, \cite{Petrecca24} found evidence of a universal shape for the power spectrum of quasars whereby, for a given accretion rate and wavelength, the PSD of quasars shifts only in frequency, according to their BH mass. Similar results were also reported by \cite{Arevalo24}.  \cite{Petrecca24} demonstrated this idea by rescaling their light curves with the light travel time of the gravitational radius of the central BH of each source, $T_{g}$, and recalculating their power spectra. They found that a bending power law (or a straight line) can fit the rescaled power spectra (multiplied by the frequency) well for all BH masses (for \medd=0.1 and $\lambda=3000$\AA). 

\begin{figure}
   \centering
    \includegraphics[width=9cm, height=7cm]{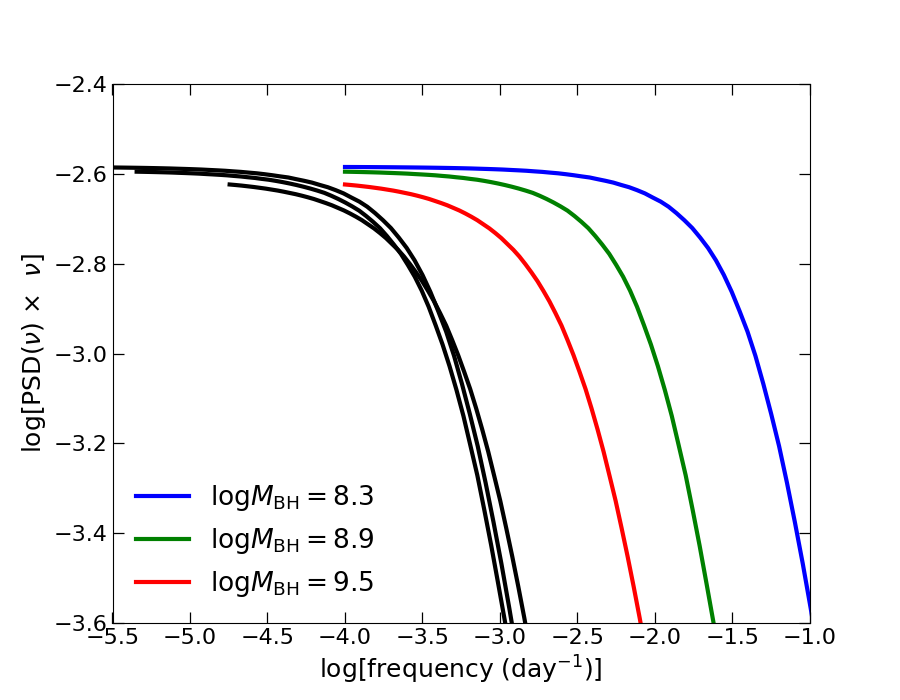}
      \caption{Model power spectrum for $\log M_{\rm BH}=8.3, 8.9, 9.5$, \medd = 0.1 and $\lambda=3000$\AA, and their rescaled versions shown with the black lines (see Sect.\ref{sec:rescaled_psd} for more details). }
         \label{fig:psd_uni}
   \end{figure}

Figure\,\ref{fig:psd_uni} shows our intrinsic model power spectra (i.e. without any aliasing correction) for the three BH masses tested here for \medd=0.1 and $\lambda=3000$\AA. By rescaling the frequencies with $T_{g}$ (i.e. dividing them by $1/T_{g}$), the models converge toward a common shape, with some scatter, suggesting that different BH masses could potentially probe different timescales of this universal PSD shape.

Our work is based on the results of \cite{Petrecca24}, who studied the quasar variability using light curves of limited duration. As a result, they managed to determine the intrinsic power spectrum of quasars over a relatively small frequency range. To significantly advance our understanding further, it is essential to obtain power spectra that span a wide frequency range, capturing variability over timescales from days to several years. In this way, key features such as the break timescale and the low- and high-frequency slopes can be reliably measured. This requires long-duration light curves (spanning many years) with dense temporal sampling of the order of a few days.

One way to accomplish this task would be to combine SDSS light curves with other existing light curves, such as the ones from Pan-STARRS-1 \citep{Chambers16}, ZTF \citep{Bellm19} and even GAIA \citep{Gaia16}. In the future, the LSST of Rubin Observatory \citep{Ivezic19} and ULTRASAT \citep{Sagiv14} will be well suited to provide the needed light curves. LSST will operate on a 10-year baseline. To compute the PSDs as accurately as possible, these 10-year-long light curves should be sampled as regularly as possible, with a dense cadence of the order of a few days (at least). 
ULTRASAT could complement these long-term light curves with minute-scale cadence ultraviolet observations over months to years (for carefully chosen samples, taking into account the differences in the rest-frame wavelength of large samples of AGNs with various redshifts). Furthermore, UVEX \citep{Kulkarni21} can serve as a powerful probe of quasar UV variability with supplemental datasets for photometric reverberation programmes, and QUVIK \citep{Werner24} can provide intense high-cadence monitoring of selected sources \citep{Zajacek24}. Together, these missions can deliver the long, well-sampled light curves on long and short timescales that are needed to construct AGN power spectra, over a broad frequency range, paving the way for robust constraints on the structure of the accretion disc and the properties of BHs.

Our work demonstrates a physical approach to studying the variability of AGNs with these upcoming surveys. Thanks to the advanced capabilities of these surveys, such an analysis will allow us to probe the accretion disc variability in a large sample of AGNs with different physical properties and across a range of redshifts and timescales, enabling a more detailed investigation of the corona's heating mechanism and the spin distribution of BHs.

\begin{acknowledgements}
We thank the referee for their useful comments that helped us improve the manuscript. This work makes use of Matplotlib \citep{Hunter07}, NumPy \citep{Harris20}, and SciPy \citep{Virtanen20}.
\end{acknowledgements}

\begin{center}
    \textbf{ORCID iDs}
\end{center}
\noindent
Marios Papoutsis \orcidlink{\orcidauthorA} \href{https://orcid.org/0009-0009-8988-0537}{https://orcid.org/0009-0009-8988-0537}\\
Iossif Papadakis \orcidlink{\orcidauthorB} \href{https://orcid.org/0000-0001-6264-140X}{https://orcid.org/0000-0001-6264-140X}\\
Christos Panagiotou \orcidlink{\orcidauthorD} \href{https://orcid.org/0009-0001-9034-6261}{https://orcid.org/0009-0001-9034-6261}\\
Elias Kammoun \orcidlink{\orcidauthorD} \href{https://orcid.org/0000-0002-0273-218X}{https://orcid.org/0000-0002-0273-218X} \\
Michal Dovčiak \orcidlink{\orcidauthorC} \href{https://orcid.org/0000-0003-0079-1239}{https://orcid.org/0000-0003-0079-1239}\\

%
%

\bibliographystyle{aa}
\bibliography{main}

\begin{appendix}

\section{The best-fit residuals}


\begin{figure}[H]
   \centering
    \includegraphics[width=9cm, height=18cm]{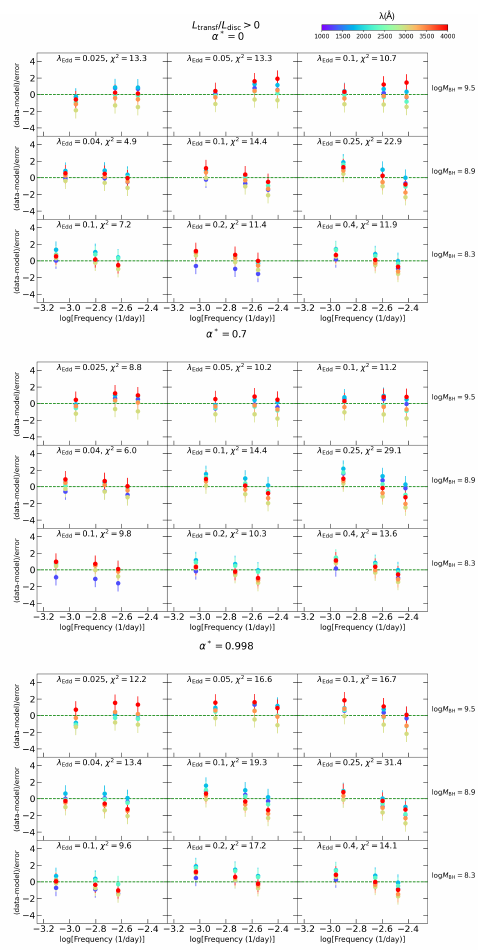}
      \caption{The best-fit residuals (i.e. (data-model)/error) for the case where the corona is powered from the accretion power of the disc (\ltransf>0). The top, middle, and bottom panels correspond to the BH spin cases \spin=0, 0.7, 0.998, respectively. The plots in the left, middle and right columns of each panel correspond to the different accretion rate cases, while the plots in the bottom, middle, and top rows to the BH masses log$M_{\rm BH}=8.3, 8.9, 9.5$ respectively. The different coloured points indicate the residuals of the fits to the 6 wavelengths of Table\,\ref{tab:parameters}.}
         \label{fig:residuals}
\end{figure}
   
\end{appendix}
\label{LastPage}
\end{document}